\documentclass[11pt, preprint]{aastex}
\usepackage{subfigure}
\let\tablehead\undefined
\let\tabletail\undefined
\usepackage{supertabular}
\let\affil\undefined
\usepackage{authblk}
\title{\bf A Second Giant Planet in 3:2 Mean-Motion Resonance in the HD 204313 System}
\author[1]{Paul Robertson}
\author[2]{J. Horner}
\author[2]{Robert A. Wittenmyer}
\author[1]{Michael Endl}
\author[1]{William D. Cochran}
\author[1]{Phillip J. MacQueen}
\author[1]{Erik J. Brugamyer}
\author[3,4]{Attila E. Simon}
\author[1]{Stuart I. Barnes}
\author[1]{Caroline Caldwell}
\affil[1]{Department of Astronomy and McDonald Observatory, University of Texas at Austin, Austin, TX 78712, USA; paul@astro.as.utexas.edu}
\affil[2]{Department of Astrophysics and Optics, School of Physics, University of New South Wales, Sydney, NSW 2052, Australia}
\affil[3]{Konkoly Observatory of the Hungarian Academy of Sciences, PO. Box 67, H-1525 Budapest, Hungary}
\affil[4]{Department of Experimental Physics and Astronomical Observatory, University of Szeged, 6720 Szeged, Hungary}

\begin{abstract}

We present 8 years of high-precision radial velocity (RV) data for HD 204313 from the 2.7 m Harlan J. Smith Telescope at McDonald Observatory.  The star is known to have a giant planet ($M \sin i = 3.5 M_J$) on a  $\sim 1900$-day orbit, and a Neptune-mass planet at $0.2$ AU.  Using our own data in combination with the published CORALIE RVs of S\'{e}gransan et al. (2010), we discover an outer Jovian ($M \sin i = 1.6 M_J$) planet with $P \sim 2800$ days.  Our orbital fit suggests the planets are in a 3:2 mean motion resonance, which would potentially affect their stability.  We perform a detailed stability analysis, and verify the planets must be in resonance.
\end{abstract}

\begin{document}

\section{\bf Introduction}

HD 204313, a sun-like $V = 8$ star observable in both hemispheres, has been a target for multiple RV surveys.  \citet{segransan10} announced the detection of the first member of the star's planetary system with the discovery of HD 204313b, a Jovian-class ($M \sin i \sim 4 M_J$) planet on a long-period ($P \sim 5$ yr) orbit.  More recently, the HARPS survey revealed an interior Neptune-mass planet with $P = 35$ days \citep{mayor11}.

At [Fe/H] = 0.18 \citep[as measured by][]{segransan10}, HD 204313 follows the observed trend of gas giant hosts being generally metal-rich \citep[e.g.][]{fischer05}.  Furthermore, planet c adds to the mounting evidence that Neptune- and lower-mass planets are extremely common around main-sequence stars \citep{howard10,mayor11,wolfgang11,witt11}.  In many ways, HD 204313 represents a ``typical'' planetary system, according to current observations.

Since 1987, we have used the 2.7 m Harlan J. Smith Telescope at McDonald Observatory for a long-baseline RV planet survey \citep{cochran93}.  An upgrade to our 2D coud\'{e} spectrograph in 1998 gave us access to the full optical wavelength range of our I$_2$ absorption cell, enabling us to monitor hundreds of FGK stars with $\sim 6$ m/s precision over 7-13 years.  One of the primary scientific objectives of the survey is to obtain a census of Jupiter analogs--giant planets in long period orbits \citep[see][for a complete discussion of Jupiter analogs and early detection limits from the McDonald Observatory RV survey]{witt06,witt11a}.  We have recently announced three giant planets in long-period orbits \citep{robertson12}, demonstrating that we have the time baseline and sensitivity to detect long-period giants.  In the core-accretion theory of giant planet formation \citep{pollack96, lissauer95}, surface-density enhancement by ices facilitates the formation of $\sim$10-15 M$_{\oplus}$ cores.  The ice line, beyond which ices are present in the protoplanetary disk, has been estimated to lie at 1.6-1.8~AU in a minimum-mass solar nebula \citep{lecar06}.  For the case of HD 204313, the inclusion of published CORALIE velocities from \citet{segransan10} gives us a total time baseline of 12 years, extending our sensitivity comfortably beyond the ice line, into the formation locations of gas giant planets.  In this paper, we present HD 204313d, another Jupiter analog exterior to planet b, and describe its orbital parameters and evolution.

\section{\bf Observations and Data Reduction}

Our RV data for HD 204313 are all taken from the 2.7 m Smith telescope between July 2003 and June 2011, resulting in an 8-year time baseline.  We use the Tull Coud\'{e} Spectrograph \citep{tull95} with a $1.8 \arcsec$ slit, yielding a resolving power R = 60,000.  Our RV measurement procedure and reduction code AUSTRAL is discussed in detail in \citet{endl00}.  In short, immediately before starlight enters the slit, it passes through an I$_2$ absorption cell regulated at 50$^{\circ}$ C, which superimposes thousands of molecular absorption lines over the object spectra in the spectral region between 5000 and 6400 \AA.  Using these lines as a wavelength standard, we simultaneously model the time-variant instrumental profile and Doppler shift relative to an I$_2$-free template spectrum.  The resulting RVs are corrected for the motion of the observatory around the solar system barycenter.  We report our RV data for HD 204313 in Table \ref{rv}.

\section{\bf Stellar Characterization}

We seek to independently verify the stellar atmosphere parameters for HD 204313 derived by \citet{segransan10}.  Using our I$_2$-free stellar template, we measure the equivalent widths of 61 Fe I lines and 17 Fe II lines.  We feed these equivalent widths to the MOOG\footnote{available at http://www.as.utexas.edu/$\sim$chris/moog.html} local thermodynamic equilibrium (LTE) line analysis and spectral synthesis program \citep{sneden73}.  By utilizing a grid of ATLAS9 model atmospheres \citep{kurucz93}, MOOG derives heavy-element abundances to match the measured equivalent widths.  We then determine effective temperature T$_{eff}$ by removing any trends in abundances versus excitation potential (assuming excitation equilibrium), and computes microturbulent velocity $\xi$ by eliminating trends with reduced equivalent width ($\equiv$ W$_{\lambda}$/$\lambda$).  Stellar surface gravity is obtained by forcing the abundances measured with Fe I and Fe II lines to match (assuming ionization equilibrium).  Our measured abundances are differential with respect to the sun.  Using a solar port, we have taken a solar spectrum using the same instrumental setup used for our RV observations, and run the above analysis for the sun.  For reference, we obtain values of T$_{eff} = 5780 \pm 70$ K, $\log g = 4.50 \pm 0.09$ dex, $\xi = 1.16 \pm 0.06$ km/s, and $\log \epsilon ($Fe$) = 7.52 \pm 0.05$ dex for the sun.  Full details of our stellar analysis can be found in \citet{brugamyer11}.

Our stellar parameters for HD 204313 are given in Table \ref{stellar}.  For values our routine does not calculate, we include catalog values from \citet[][Version 3]{kharchenko09} and \citet{casagrande11}.  Our computed values agree extremely well with those presented in \citet{segransan10}.  Of particular interest is our measured [Fe/H] of $0.24 \pm 0.06$, which confirms the metal-rich nature of HD 204313.  

\section{\bf Orbit Modeling}

\begin{figure}
  \begin{center}
    \subfigure[\label{rvplot}]{\includegraphics[scale=0.3]{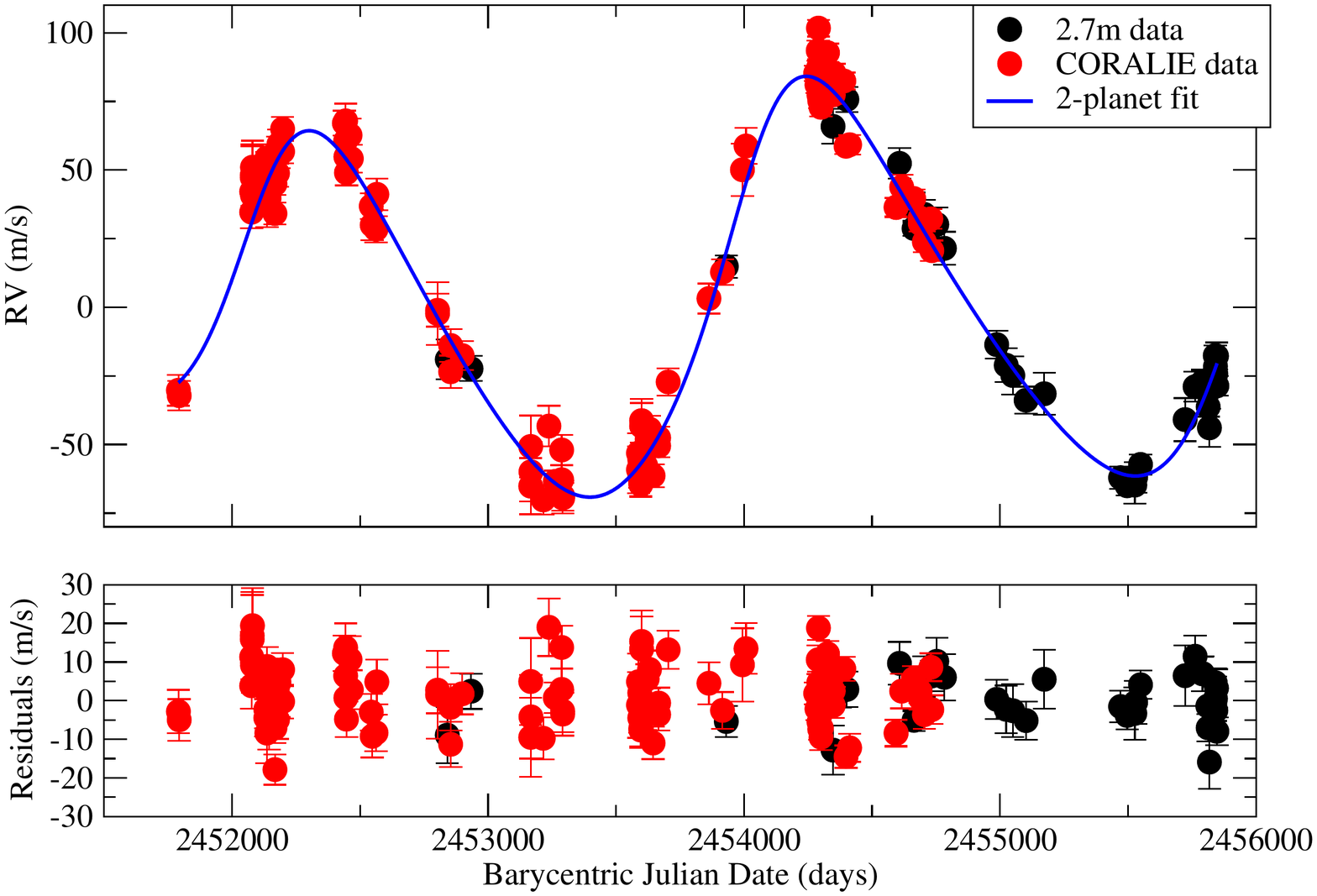}}
    \subfigure[\label{bplot}]{\includegraphics[scale=0.3]{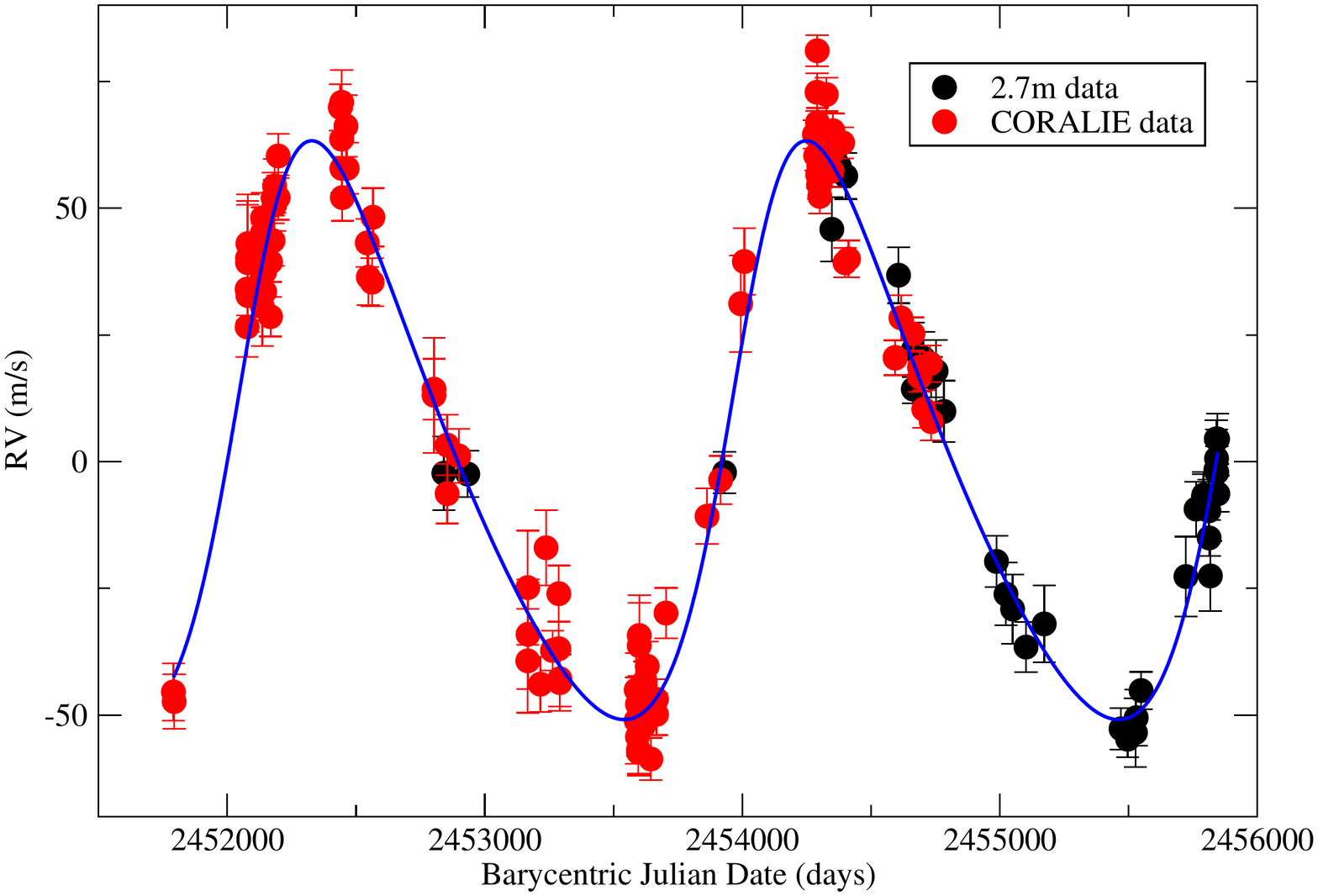}}
    \subfigure[\label{dplot}]{\includegraphics[scale=0.3]{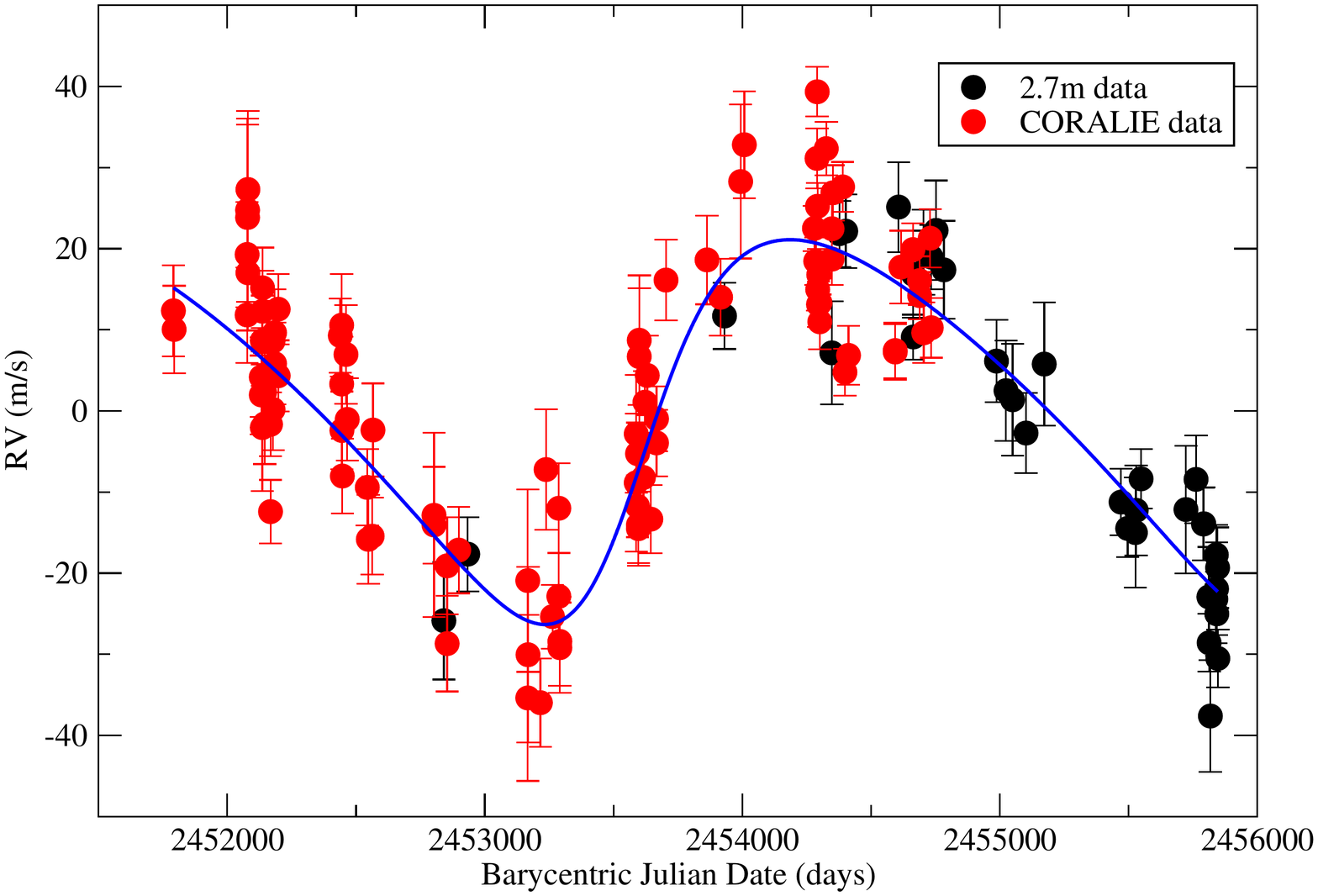}}
    \caption{a.  \emph{Top}: Radial velocity data for HD 204313.  Points in black are our 2.7m observations, while points in red are CORALIE observations from \citet{segransan10}.  The best-fit orbit model is shown as a blue line.  \emph{Bottom}: Residuals to a two-planet fit.  b.  RVs after subtracting our fit to planet d from the velocities in (a).  The blue line shows our Keplerian model for planet b.  c.  RVs after subtracting our fit to planet b from the velocities in (a).  The blue line shows our Keplerian model for planet d.}
    \end{center}
\end{figure}

Over the 8-year period from July 2003 to June 2011, we have collected 36 RV points for HD 204313.  Our data are plotted as a time series in Figure \ref{rvplot}.  The RMS scatter about the mean of these velocities is 40 m/s, with an average internal error of 5.21 m/s.  When analyzed alone, our RVs show the high-amplitude ($K \sim 70$ m/s) signal expected as a result of HD 204313b, with a period around 5.5 years.  We compute Keplerian orbital solutions using the GaussFit modeling program \citep{jefferys88} and the SYSTEMIC console \citep{meschiari09}, finding excellent agreement between the one-planet solutions from both routines.  However, a one-planet fit to our data gives a period more than 100 days longer than the period reported in \citet{segransan10}, a discrepancy more than 3 times the combined $1\sigma$ uncertainties in the orbital period for the two models.  Additionally, our fit includes a long-period linear trend.  We note that \citet{mayor11} also find a period for planet b considerably longer than the originally published value.

We then compute a one-planet model using the CORALIE RVs as well as our own.  The combined RV set includes 132 RVs taken over 10 years. The resulting parameters are closer to the previously published solution; we find a period of 2,000 days, with eccentricity 0.16 and a minimum mass of 4.36 Jupiter masses.  However, this fit is still considerably discrepant from the \citet{segransan10} solution, and leaves a residual RMS scatter of 11.0 m/s, and a reduced $\chi^2 = 5.66$.  

\begin{figure}
\begin{center}
\includegraphics[scale=0.7]{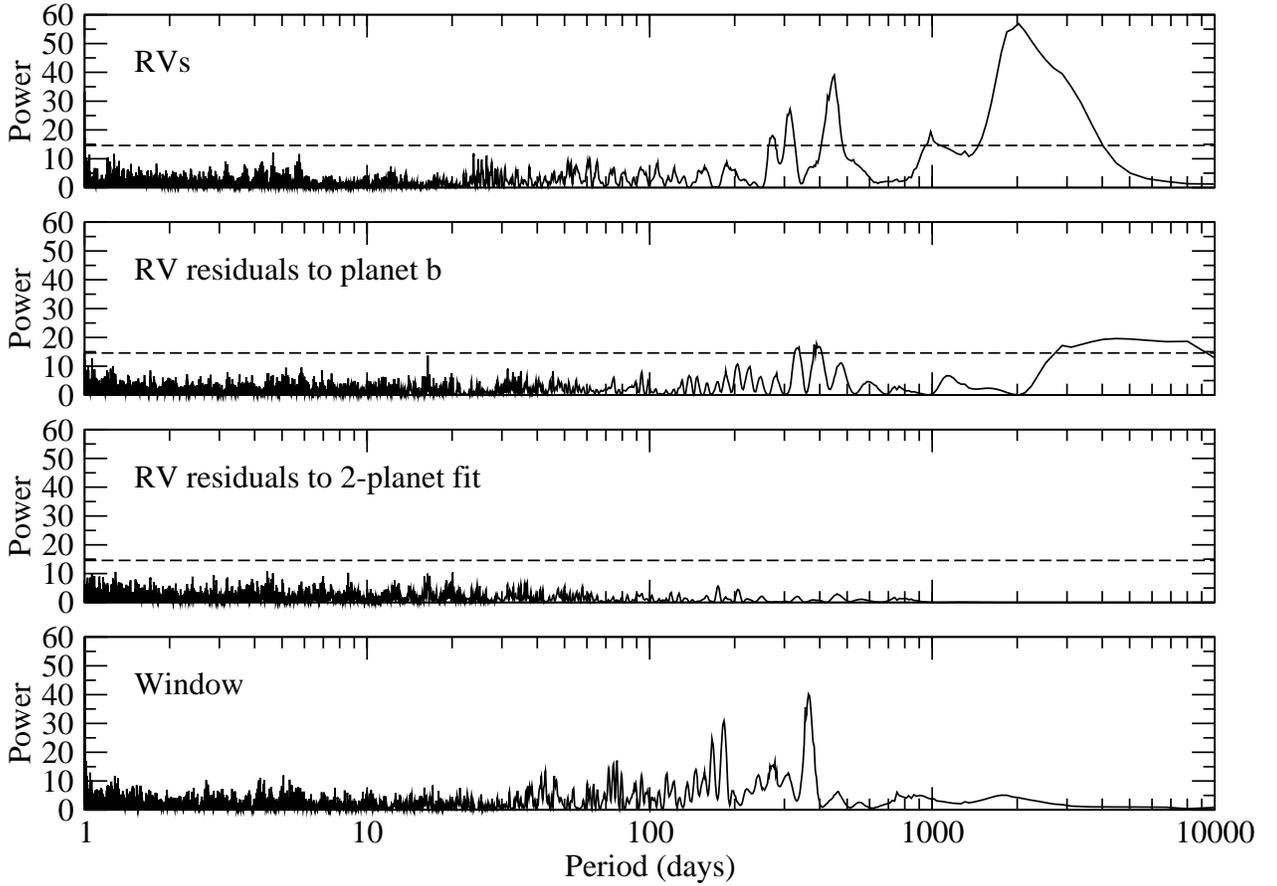}
\caption{\emph{From top}: [\emph{1}] Generalized Lomb-Scargle periodogram for the combined CORALIE/McDonald RVs of HD 204313.  [\emph{2}] The same periodogram for the residual RVs after subtracting a one-planet fit.  [\emph{3}] Periodogram of the residual RVs after subtracting a two-planet fit.  [\emph{4}] Periodogram of our time sampling (the window function).  The dashed lines indicate the approximate power level for a FAP of 0.01, computed from Equation 24 of \citet{zechmeister09}.}
\label{ps}
\end{center}
\end{figure}

We have computed the fully generalized Lomb-Scargle periodogram \citep{zechmeister09} for the combined CORALIE-McDonald data set, and the residual RVs around the one-planet fit.  The resulting power spectrum is shown in Figure \ref{ps}.  The periodogram of the residual RVs shows significant peaks around 340 days, 395 days, and a broad peak between 2700 and 6700 days.  We calculate a false-alarm probability (FAP) for these peaks using the method described in \citet{sturrock10}, and find a FAP of approximately $5 \times 10^{-5}$ for the long-period peak, while the 395-day and 340-day peaks have FAPs of $4 \times 10^{-4}$ and $1.5 \times 10^{-3}$, respectively.

We attempt to fit an additional planet to the residuals at each of the periods identified in the periodogram.  Our fitting routine produces unsatisfactory solutions at the two shorter periods, but converges to a fit with an outer giant ($M \sin i = 1.68 M_J$) planet at 2831 days.  The period of planet b in this solution is consistent with the original published result, although the eccentricity is higher than in a one-planet fit.  The resultant two-planet fit is included in Figure \ref{rvplot}, and we give residual plots of the individual planets in Figures \ref{bplot} and \ref{dplot}.  As with the one-planet solution, GaussFit and SYSTEMIC agree nicely on the orbital parameters and their uncertainties.  The parameters of our final orbital model are listed in Table \ref{orbit}.  The addition of planet d removes the need to include a linear slope.  Although we attempted to fit an outer planet to each of the CORALIE and McDonald RV sets individually, our routines failed to converge for either set.  Evidently, both data sets are required to achieve the time baseline needed to detect planet d, a fact reinforced by our periodogram analysis.  When examining the residual RVs to our one-planet fit for each data set individually, we see only a monotonic increase in power at long periods for the CORALIE data and insignificant power in the McDonald data.  Only when the data are combined, and the total time baseline exceeds a full orbit of planet d, does the power spectrum show a clearly-defined peak around the period of that planet.
  
We note that we have not included planet c \citep{mayor11} in our analysis.  With a reported RV amplitude of just 3.28 m/s, the signal of this short-period planet is below the sensitivity limit of our data and that of CORALIE.  Indeed, our periodogram of the residuals to our two-planet solution (Figure \ref{ps}) shows no additional signals.  Furthermore, the inclusion of a third planet with the orbital elements published for planet c does not significantly change our orbital solution.  However, the RMS (7.79 m/s) and reduced $\chi^2$ (2.98) of our two-planet model are still higher than we expect given the precision of the Tull spectrograph and the CORALIE data.  While this is reflected in our model as a relatively high level of stellar ``jitter'' (5.46 m/s), our stellar activity analysis (see below) suggests HD 204313 should not be so active.  Although it would be ideal to include the orbit of planet c in our model to verify this hypothesis, we are unable to do so because \citet{mayor11} include neither their measured RVs nor their complete orbital fit for the HD 204313 system.  We nevertheless conclude that the additional scatter around our fit is most likely due to the unresolved planet c, and potentially additional low-mass companions.  We refer to the outer planet as HD 204313d in acknowledgement of the inner Neptune-mass planet.

\citet{eggenberger07} report a companion star $6.2\arcsec$ to the south of HD 204313, although they admit a significant probability of a chance alignment.  At a distance of 47 pc, the angular separation indicates a minimum physical distance of 583 AU between the two objects.  The companion is approximately 9 magnitudes fainter in the near infrared \citep{eggenberger07}, and is therefore much less massive than HD 204313 if they are in fact bound.  If we overestimate the mass of this object at $0.5 M_{\odot}$ and assume it is associated with HD 204313, the resulting radial velocity slope due to the companion is 0.28 m s$^{-1}$ yr$^{-1}$, which is roughly equal to our $1 \sigma$ uncertainty level of 0.2 m s$^{-1}$ yr$^{-1}$ for a slope in the combined data set.  It is therefore safe to conclude that the second star is not influencing our modeling of the planetary system.

\section{\bf Stellar Activity and Line Bisector Analysis}

While we do not anticipate that stellar activity should produce RV signals of the amplitudes of planets b and d, it is nevertheless important to understand how changes in the atmosphere of HD 204313 may influence our velocity measurements, particularly with the amount of scatter seen around our fit.  We examine stellar activity simultaneously with RV through line bisector analysis of stellar lines outside the I$_2$ region and changes in the Ca H and K indices.

Changes in the stellar photosphere (starspots, etc.) may produce changes in the measured RVs.  However, these processes will also alter the shapes of the individual stellar absorption lines.  Following the method of \citet{brown08}, we calculate the bisector velocity span (BVS) for each of our spectra.  The BVS is sensitive to these subtle changes in line shapes, and therefore a reliable indicator of activity the stellar photosphere.

Similarly, if stellar activity is producing RV signals, those signals should also appear in the Ca H and K indices.  For each RV point in Table \ref{rv}, we have computed the Mount Wilson $S_{HK}$ index, which we list alongside the velocities.  From \citet{noyes84} we use $S_{HK}$ to derive $\log R'_{HK}$, the ratio of Ca H and K emission to the bolometric luminosity of the star.  From $\log R'_{HK}$ we obtain a more general idea of the overall activity level of HD 204313.

\begin{figure}
\begin{center}
\subfigure[\label{bvs}]{\includegraphics[scale=0.5]{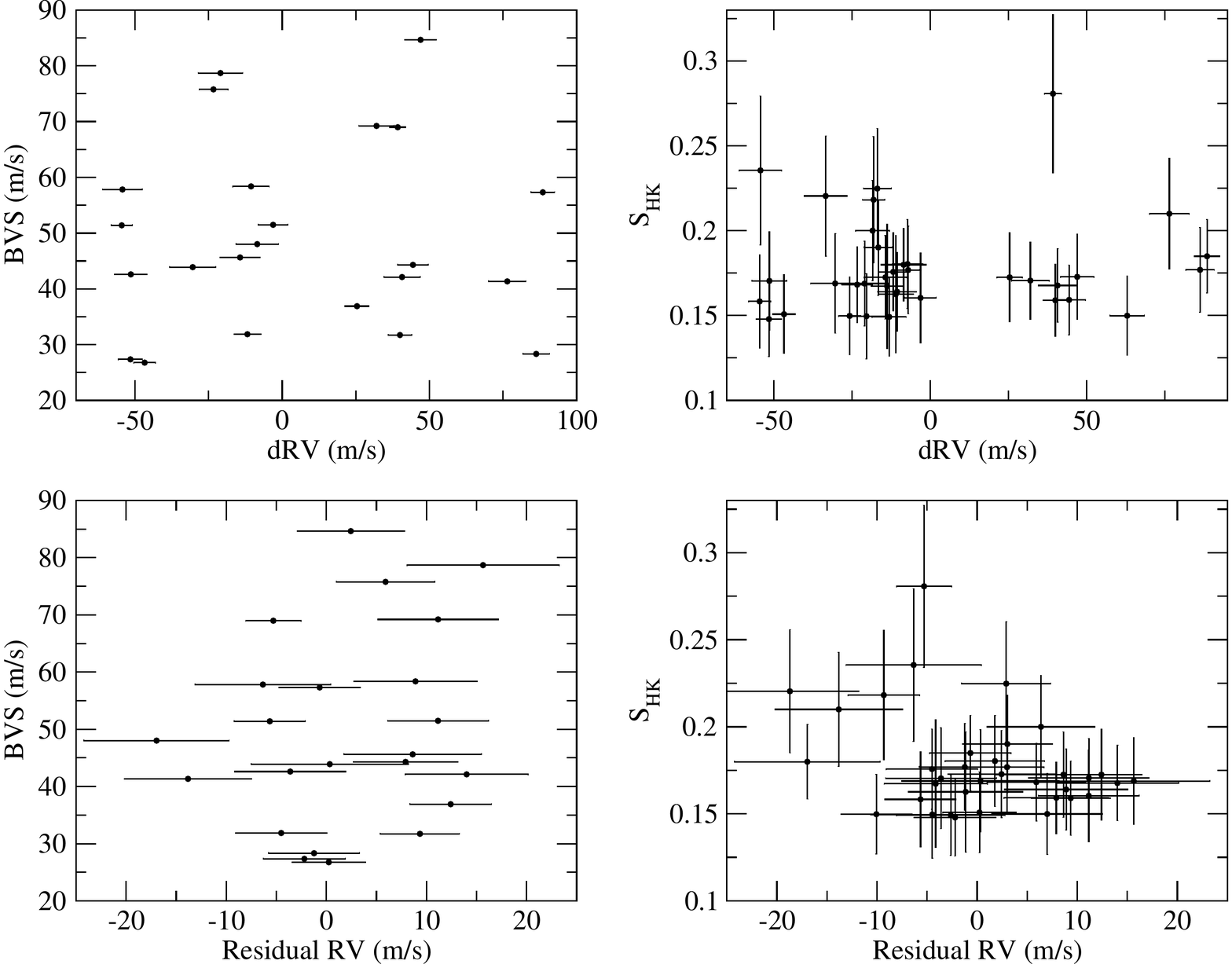}}
\subfigure[\label{bvsps}]{\includegraphics[scale=0.4]{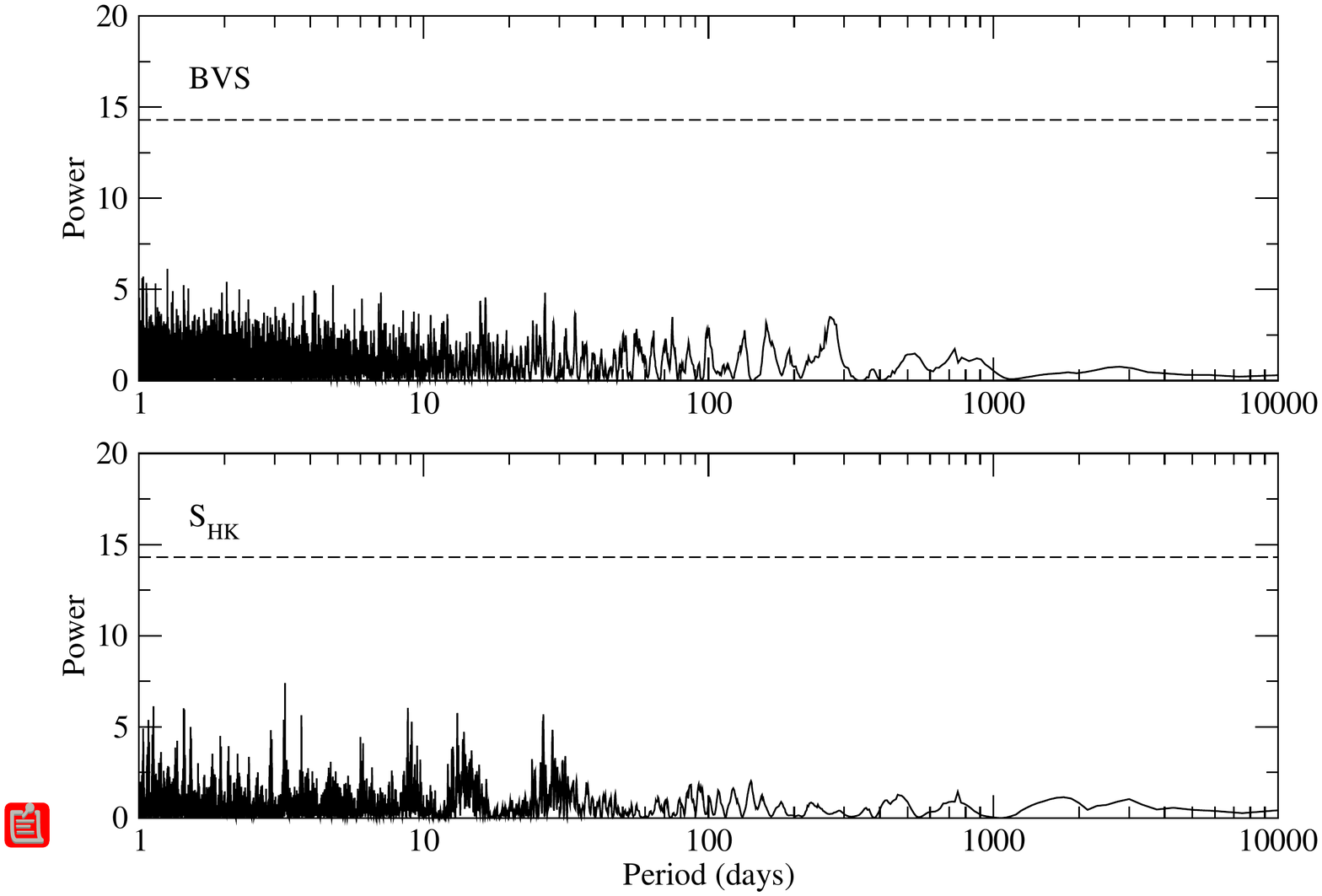}}
\caption{a.  \emph{Left}: Bisector velocity spans plotted against our measured RVs (\emph{top}) and residual RVs to a one-planet fit (\emph{bottom}) for HD 204313.  \emph{Right}: $S_{HK}$ indices plotted against our measured RVs (\emph{top}) and residual RVs to a one-planet fit (\emph{bottom}) for HD 204313.
\newline
\newline
b.  Generalized Lomb-Scargle periodograms for the BVS (\emph{top}) and $S_{HK}$ indices (\emph{bottom}) of our spectra for HD 204313.  The dashed lines indicate the approximate power level for a FAP of 0.01.}
\label{activity}
\end{center}
\end{figure}

All examinations show HD 204313 to be an extremely quiet star.  The results of our activity analyses are shown in Figure \ref{activity}.  In Figure \ref{bvs}, we plot BVS and $S_{HK}$ versus our measured RVs and their residuals around the one-planet fit.  In both cases, there is no significant correlation, suggesting photospheric activity is not influencing our velocities.  Periodograms of $S_{HK}$ and BVS (Figure \ref{bvsps}) show no periodicity for either index.  Furthermore, we measure an RMS of only 17 m/s for the BVS, and $\log R'_{HK} = -4.65$.  It is not surprising, then, that we see no signals or correlations in any of our activity indicators.  With three planets now known, HD 204313 is rapidly becoming a rich planetary system.  Its low activity level makes it an ideal candidate for follow-up observations to search for additional low-mass companions.

\section{\bf Dynamical Stability Analysis}

A number of recent studies have highlighted the need for observational detections of multiple exoplanet systems to be supported by dynamical simulations that test whether the orbits of the proposed planets are dynamically feasible \citep[e.g.][]{horner11,horner12,witt12,hinse12}. Such studies are particularly important when the planets in question appear to move on orbits close to mutual mean-motion resonance \citep[e.g.][]{robertson12}, an architecture that can yield either extreme stability or instability, depending on the precise orbits of the planets involved. In the case of HD 204313, the best-fit orbits for planets b and d suggest that they may well be trapped in mutual 3:2 mean-motion resonance. As such, we chose to perform a highly detailed dynamical study of the orbits of planets b and d to investigate whether the orbits that best fit the data are dynamically feasible. 

Following earlier work \citep{horner11,marshall10,robertson12}, we used the \emph{Hybrid} integrator within the $n$-body dynamics package MERCURY \citep{chambers99} to examine test systems in which the initial orbit of planet b was held fixed at the nominal best fit values (in this case, $a = 3.04$ AU, $e = 0.23$). The initial orbit of planet d was then systematically changed from one simulation to the next, such that scenarios were tested for orbits spanning the full $\pm 3\sigma$ error ranges in semi-major axis, eccentricity, longitude of periastron and mean anomaly. Such tests have already proven critical in confirming or rejecting planets thought to follow unusual orbits \citep[e.g.][]{horner11,witt12}, and allow the construction of detailed dynamical maps for the planetary system studied, in orbital element phase space.

We examined 31 unique values of semi-major axis for planet d, ranging from 3.51 AU to 4.35 AU, inclusive, in even steps. For each of these 31 initial semi-major axes, we studied 31 values of orbital eccentricity, ranging across the full $\pm 3\sigma$ range ($e = 0.01-0.55$). For each of the resulting 961 $a$-$e$ pairs, we considered 11 values of initial longitude of periastron ($\omega$), and 5 values of initial mean anomaly ($M_0$), resulting in a total suite of 52,855 ($31 \times 31 \times 11 \times 5$) plausible architectures for the HD 204313 system.

In each of these simulations, the masses of the two planets studied were set to their minimum ($M \sin i$) values. The mass of planet b was therefore set to $3.55 M_J$, while that of planet d was set to $1.68 M_J$. To first order, the more massive the planets, the more strongly they will perturb one another, and so setting their masses to the minimum allows us to maximize the potential stability of the planetary system. In other words, we expect our resulting dynamical maps to show the maximal stability of the orbits tested. The dynamical evolution of the two planets was then followed for a period of 100 million years, or until one of the planets either collided with the central star, was transferred to an orbit that took it to a distance of at least 10 AU from the central star, or collided with the other planet.  Collisions were modeled by assuming a density of 1.33 g cm$^{-3}$--equal to the average density of Jupiter--for each planet and computing a radius accordingly, so that our code registered a collision if an actual physical encounter occured.  The time of such events was recorded, allowing us to construct a dynamical map of the planetary system, shown in Figure \ref{meanae}. That figure shows the mean lifetime of the HD 204313b-d system as a function of the initial semi-major axis, $a$, and eccentricity, $e$, of planet d. Each individual initial $a$-$e$ pair was tested a total of 55 times, each of which featured a different initial combination of $\omega$-$M_0$. The lifetimes shown are the mean value of the 55 individual lifetimes obtained from those runs. 

Aside from resonant solutions, the entire $a$-$e$ phase space of allowed orbits for planet d is extremely unstable, with collisions between planets b and d often occurring within the first few hundred years of the simulations. It is also clear that, even within the resonance, some subset of the solutions are dynamically unstable (hence the reason the mean lifetime in the stable region is somewhat less than $10^8$ years). At the highest eccentricities permitted for the orbit of planet d, no stable solutions exist, but there is a broad region of stability within the $1\sigma$ errors on the best-fit orbit. As can be seen in Figure \ref{meanaw}, the stability of orbits in the vicinity of the 3:2 mutual mean-motion resonance between the planets is a strong function of the longitude of periastron $\omega$ for planet d (we note here that planet b's initial longitude of periastron was 298 degrees).  Qualitatively, the strong $\omega$ dependence is reflective of the fact that the 3:2 resonance provides stability by ensuring planets b and d never simultaneously approach a true anomaly $\nu \sim 300^{\circ}$, where their orbital paths allow very small separations.  For configurations outside the stable $a-\omega$ space, the resonance becomes destructive.  Once again, the stable region extends throughout the $1\sigma$ uncertainties on the best-fit orbit of planet d, with the location of the best-fit orbit lying close to the region of greatest stability. 

Here again, our solution suffers from a lack of information regarding planet c.  Fortunately, at $P = 35$ days and $M \sin i = 17 M_{\oplus}$, plausibility arguments suffice to rule out destabilizing interactions due to this inner planet.  As demonstrated in \citet{horner11}, planets tend to be stable when separated by $\sim 5$ Hill radii.  With planets b and c separated by nearly 12 Hill radii (measured from planet b), we expect little mutual influence.  A similar argument is presented for KOI 961 \citep{muirhead12}.  Nevertheless, we have performed a small number of simulations in which we include a planet with $M = 17 M_{\oplus}$, $a = 0.21$ AU, and $e = 0.17$ \citep{mayor11} to the stable configurations nearest to our best-fit orbital solution.  In all cases, the stability of the system is unaffected; while the exact values of $a$ and $e$ for the giant planets are slightly different at each time step when planet c is included, the periods over which the orbital parameters vary remain unchanged, and the long-term evolution is the same regardless of whether c is included.  We therefore conclude that excluding planet c from our larger analysis does not significantly affect our results.

Taken in concert, the results shown in Figure \ref{stability} reveal that dynamically stable coplanar solutions for the orbit of planet d require that it be trapped in mutual mean-motion resonance with planet b. Given that the nominal best-fit orbit lies perfectly within the region spanned by that resonance, and that a significant fraction of the $1\sigma$ error ellipse for planet d is dynamically stable in both $a$-$e$ and $a$-$\omega$ space, we find that our dynamical results are broadly in support of the existence of planet d, and may even be used to more tightly constrain its orbit.

\section{\bf Discussion}

With the addition of planet d, HD 204313 joins the growing list of stars hosting multiple gas giant planets.  At G5 V spectral type and at virtually equal mass to the sun, having two planets with masses $\sim$ 2-4 times that of Jupiter makes HD 204313 somewhat of an outlier on the correlation between stellar mass and giant planet fraction/mass \citep{johnson11}.  However, we confirm the star's super-solar metallicity measured by \citet{segransan10}, thereby offsetting the slight discrepancy with stellar mass.

As of May 2012, there are 12 exoplanet systems in the exoplanets.org \citep{wright11} database which contain giant planets believed to be in low-order resonances.  However, only HD 45364 \citep[3:2,][]{correia09} and HD 200964 \citep[4:3,][]{johnson11b} host multiple gas giants in mean-motion resonances closer than 2:1.  HD 204313 therefore joins a very small subset of the known planet systems.  Furthermore, the minimum masses of planets b and d are considerably higher than either the HD 45364 or HD 200964 planets, making their continued stability even more remarkable.  Interestingly, in addition to having the 3:2 resonance in common, HD 45364b/c and HD 204313b/d both have mass ratios close to the $\sim$ 3:1 Jupiter/Saturn mass ratio.  These systems are thus valuable as a comparison to the Nice model \citep{tsiganis05} for the formation of the outer solar system.  In particular, \citet{batygin12} invoke a 3:2 resonance between Jupiter and Saturn in simulations which successfully reproduce the orbital configurations of the four outer solar system planets and the Kuiper belt.

\begin{figure}
\begin{center}
\subfigure[\label{meanae}]{\includegraphics[scale=0.75]{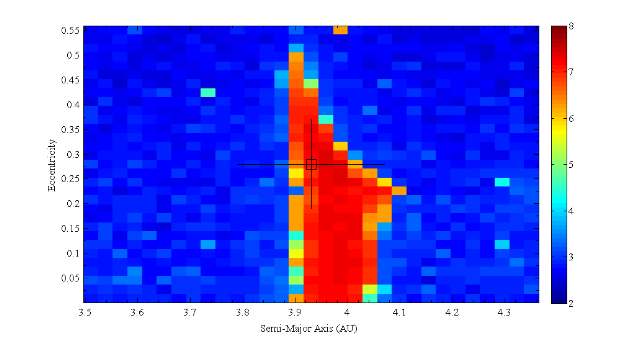}}
\subfigure[\label{meanaw}]{\includegraphics[scale=0.75]{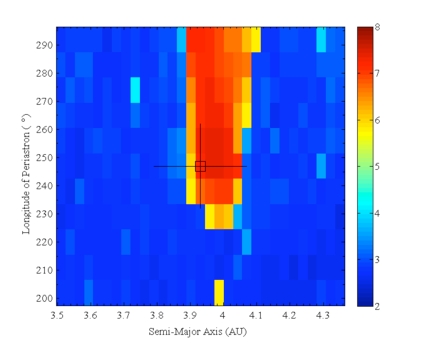}}
\caption{\emph{a}: The mean dynamical lifetime of the HD 204313b-d planetary system, as a function of the initial semi-major axis ($a$) and eccentricity ($e$) of planet d. The lifetimes are shown on a logarithmic scale, ranging from $10^2$ years (blue) to $10^8$ years (red). The location of the nominal best-fit orbit for planet d is denoted by the open square, with the $1\sigma$ uncertainties shown by the solid lines radiating from that point. It is immediately obvious that the vast majority of the $a$-$e$ space tested is highly unstable, with only a narrow region of stability centered on the mutual 3:2 mean-motion resonance between planets b and d.
\newline
\newline
\emph{b}: The mean lifetime of the HD 204313b-d system, as a function of the semi-major axis, $a$, and longitude of periastron, $\omega$, of planet d's orbit. The lifetime shown at each location in $a$-$\omega$ space is the mean of 155 individual runs, which tested 31 different orbital eccentricities and 5 different mean anomalies for that particular $a$-$\omega$ combination. The color scheme is the same as in \emph{a}, and the nominal best-fit orbit for planet d is again marked by the unfilled square, with the $1\sigma$ errors on that fit shown by the solid lines that radiate from that point.}
\label{stability}
\end{center}
\end{figure}

\begin{figure}
\begin{center}
\includegraphics[scale=0.75]{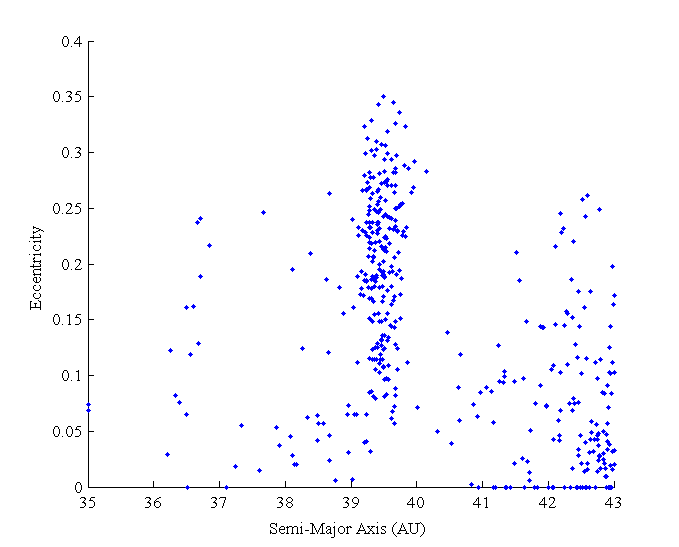}
\caption{The distribution in $a$-$e$ space of the population of trans-Neptunian objects between 35 and 43 AU. Note the concentration of objects just beyond 39 AU - the Plutinos, trapped in 3:2 mean-motion resonance with Neptune.}
\label{tno}
\end{center}
\end{figure}

The importance of 3:2 mean-motion resonances within the Solar system extends beyond the possible interactions between Jupiter and Saturn during the system's early evolution. Beyond the orbit of Neptune lie the Plutinos (named after the dwarf planet (134340) Pluto, the first known member). These objects, of which several hundred are currently known, are trapped within the 3:2 Neptunian mean-motion resonance.  In Figure \ref{tno}, we plot the Plutino distribution in $a$-$e$ space.  The Plutino population contains objects with a wide range of eccentricities and inclinations, with the most eccentric objects crossing the orbit of Neptune, and some moving on orbits that can range as close as halfway between the orbits of Uranus and Neptune. The inclinations of the Plutinos range from 0 degrees to over thirty degrees. This wide distribution of orbital elements has been used to decipher the migration history of Neptune - the idea being that, as that giant planet migrated outwards, objects were captured into the 3:2 mean-motion resonance and swept along with the planet, their orbits becoming ever more excited as they were carried along \citep[e.g.][]{malhotra95}. As a result, the Plutinos nicely map the extent of the stable region of the 3:2 mean-motion resonance with Neptune. Though the stable Plutinos do not range to quite as extreme eccentricities as are supported for the orbit of HD204313d (as a result of the influence of Uranus on the evolution of the most eccentric members), it is striking that the region of stability occupied by the Plutinos is very similar to that obtained by our dynamical integrations, as can be seen when comparing Figures \ref{meanae} and \ref{tno}. We note in passing that the Hilda family of main belt asteroids are trapped in 2:3 mean motion resonance with Jupiter, orbiting with periods of $\sim 8$ years. Despite their sometimes high orbital eccentricities (again up to, and in excess of, 0.3), these objects are protected from close encounters with the massive planet by the mean-motion resonance they occupy.

Unlike our recent results for the 2:1 mean-motion resonance in the HD 155358 system \citep{robertson12}, our dynamical simulations for HD 204313 do not permit coplanar orbits outside the 3:2 resonance.  Evidently, such a small period ratio is only stable when protected by the resonance.  Furthermore, the existence of unstable orbits within the permitted parameter space shows that location within a resonance is not a guarantee of stability.  Finally, it is interesting that while for both HD 155358 and HU Aquarii \citep{horner11,witt12}, the inner cutoff for stability in $a$-$e$ space is equal to the inner planet's apastron distance plus five Hill radii, our stable solutions for HD 204313 allow planet d to have values of $a$ much smaller than this value.  Once again, such a conclusion is supported by our knowledge of our Solar system, in which resonant configurations ensure the stability of vast populations of objects on orbits that would otherwise be highly unstable. Prominent examples include the aforementioned Hildas \citep[e.g.][]{franklin93,grav12} and Plutinos \citep{malhotra95,friedland01}, along with the Jovian and Neptunian Trojans \citep{morbidelli05,sheppard06,lykawka10,horner12b}. In addition to the resonant exoplanets mentioned earlier, these populations reinforce the idea that such resonant scenarios are a common outcome from the planet formation process.

We note that the results of our stability analysis not only confirm the validity of our orbital model, but in fact place tighter constraints on the system's configuration than our fitting uncertainties alone.  As long-term RV planet surveys such as ours become increasingly sensitive to systems with multiple long period companions, it is likely that additional systems with gas giants in close resonances will be discovered.  Such dynamical simulations are therefore extremely valuable for understanding the true architecture of these systems, for which there may otherwise be considerable uncertainty as to the orbital parameters.

It is important to note that our claim that planets b and d are trapped in mutual 3:2 mean-motion resonance is based on simulations that assumed that their orbits are coplanar. However, as can be seen from the examples of the Hildas and Plutinos within our own Solar system, resonant orbits can be dynamically stable for a wide range of mutual inclinations. It might instinctively seem that the coplanar case would actually be the least stable configuration, and therefore that mutually inclined orbits might allow a broader range of stable solutions. However, we note that in \citet{horner11}, the authors considered a wide range of orbital inclinations in an attempt to address the apparent instability of the proposed HU Aquarii planetary system, and found that increasing the mutual inclination of the planets in question did little to remedy their instability. That said, it would certainly be interesting, in future work, to examine the influence of the mutual inclination of the orbits of the planets of the HD204313 system.  Fortunately, with predicted astrometric displacements of 0.432 mas and 0.264 mas for planets b and d, respectively, both planets should be accessible to astrometric measurements with the HST Fine Guidance Sensor \citep{nelan10}.  Plus, the inclusion of inclination constraints would make HD 204313 a unique opportunity for comparison to the compact, multi-resonant planet systems discovered by the \emph{Kepler} spacecraft \citep{holman10,lissauer11,cochran11}.   If astrometry shows the HD 204313 planets to be coplanar, it would be strongly suggestive that similar migration mechanisms can result in systems as different as HD 204313 and the aforementioned \emph{Kepler} planets.  HD 204313 should therefore be considered a high-priority target for current and future astrometric surveys.

\begin{acknowledgements}
J.H. gratefully acknowledges the financial support of the Australian government through ARC Grant DP0774000.  R.W. is supported by a UNSW Vice-Chancellor's Fellowship.  M.E. and W.D.C. acknowledge support by the National Aeronautics and Space Administration under Grants NNX07AL70G and NNX09AB30G issued through the Origins of Solar Systems Program.  A. E. Simon has been supported by the Hungarian OTKA  Grants K76816, K83790 and MB08C 81013, the ``Lendulet"  Program of the Hungarian Academy of Science.  This research has made use of the Exoplanet Orbit Database and the Exoplanet Data Explorer at exoplanets.org.
\end{acknowledgements}

\clearpage

\begin{figure}
\begin{center}
\includegraphics[scale=0.7]{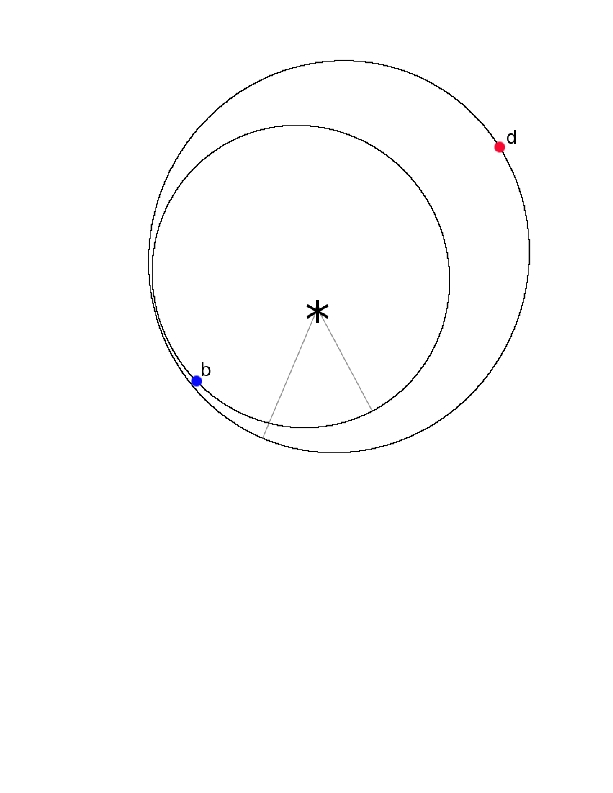}
\caption{Face-on orbital diagram of the giant planets in the HD 204313 system.  The ellipses shown are derived from the model in Table \ref{orbit} (the open square in Figure \ref{stability}), with the lines from the star pointing toward the periastron of each planet.  The locations of planets b and d are adopted from the mean anomalies in Table \ref{orbit}.}
\label{orbitplot}
\end{center}
\end{figure}

\clearpage

\begin{table}
\begin{center}

\footnotesize

\tablecaption{Radial Velocities for HD 204313}
\label{rv}	
\tablefirsthead{\hline
BJD - 2450000 & Radial Velocity & Uncertainty & $S_{MW}$ \\
& (m/s) & (m/s) & \\
\hline}
		
\tablehead{\hline
\emph{Table \label{rv} cont'd.} & & & \\ \hline
BJD - 2450000 & Radial Velocity & Uncertainty &$S_{MW}$ \\
& (m/s) & (m/s) & \\
\hline}
		
\tabletail{\hline}

\begin{supertabular}{| l l l l |}

2840.892291 	 & -8.48 	 & 7.26 	 & $ 0.1799 \pm 0.0215 $ \\
2933.666733 	 & -11.81 	 & 4.58 	 & $ 0.1757 \pm 0.0231 $ \\
3930.951397 	 & 25.41 	 & 4.08 	 & $ 0.1725 \pm 0.0262 $ \\
4347.740522 	 & 76.44 	 & 6.37 	 & $ 0.2100 \pm 0.0327 $ \\
4376.744143 	 & 88.50 	 & 4.07 	 & $ 0.1849 \pm 0.0217 $ \\
4401.617979 	 & 86.26 	 & 4.55 	 & $ 0.1769 \pm 0.0250 $ \\
4606.945812 	 & 62.97 	 & 5.54 	 & $ 0.1499 \pm 0.0233 $ \\
4663.924822 	 & 46.99 	 & 5.37 	 & $ 0.1728 \pm 0.0251 $ \\
4663.938000 	 & 39.25 	 & 2.76 	 & $ 0.2807 \pm 0.0466 $ \\
4703.749028 	 & 44.40 	 & 5.24 	 & $ 0.1592 \pm 0.0206 $ \\
4732.798800 	 & 40.00 	 & 3.96 	 & $ 0.1590 \pm 0.0214 $ \\
4752.702889 	 & 40.71 	 & 6.14 	 & $ 0.1677 \pm 0.0217 $ \\
4782.617353 	 & 32.02 	 & 6.04 	 & $ 0.1706 \pm 0.0228 $ \\
4986.953099 	 & -3.09 	 & 5.05 	 & $ 0.1604 \pm 0.0265 $ \\
5023.896398 	 & -10.58 	 & 6.18 	 & $ 0.1640 \pm 0.0233 $ \\
5049.798314 	 & -14.30 	 & 6.87 	 & $ 0.1725 \pm 0.0246 $ \\
5101.700768 	 & -23.34 	 & 4.92 	 & $ 0.1682 \pm 0.0223 $ \\
5172.538410 	 & -20.96 	 & 7.60 	 & $ 0.1688 \pm 0.0250 $ \\
5470.683390 	 & -51.54 	 & 4.11 	 & $ 0.1479 \pm 0.0222 $ \\
5496.568795 	 & -54.50 	 & 3.56 	 & $ 0.1583 \pm 0.0275 $ \\
5526.556625 	 & -54.25 	 & 6.77 	 & $ 0.2355 \pm 0.0439 $ \\
5529.556590 	 & -51.40 	 & 5.55 	 & $ 0.1704 \pm 0.0290 $ \\
5548.544742 	 & -46.72 	 & 3.67 	 & $ 0.1508 \pm 0.0232 $ \\
5722.947515 	 & -30.38 	 & 7.86 	 & $ 0.1689 \pm 0.0294 $ \\
5761.927802 	 & -18.40 	 & 5.41 	 & $ 0.2000 \pm 0.0296 $ \\
5790.814330 	 & -16.89 	 & 4.45 	 & $ 0.2248 \pm 0.0355 $ \\
5791.852748 	 & -16.59 	 & 4.50 	 & $ 0.1901 \pm 0.0282 $ \\
5811.740128 	 & -20.36 	 & 6.16 	 & $ 0.1496 \pm 0.0251 $ \\
5812.750735 	 & -25.75 	 & 3.56 	 & $ 0.1498 \pm 0.0229 $ \\
5817.780745 	 & -33.40 	 & 6.91 	 & $ 0.2204 \pm 0.0354 $ \\
5838.671924 	 & -13.10 	 & 5.43 	 & $ 0.1492 \pm 0.0231 $ \\
5840.686936 	 & -7.06 	 & 3.68 	 & $ 0.1768 \pm 0.0259 $ \\
5841.664865 	 & -10.98 	 & 5.72 	 & $ 0.1626 \pm 0.0346 $ \\
5842.697923 	 & -13.76 	 & 5.16 	 & $ 0.1673 \pm 0.0367 $ \\
5845.589786 	 & -7.23 	 & 4.97 	 & $ 0.1803 \pm 0.0262 $ \\
5846.724380 	 & -18.09 	 & 3.57 	 & $ 0.2182 \pm 0.0372 $ \\

\end{supertabular}

\end{center}
\end{table}

\begin{table}
\begin{center}

\begin{tabular}{| l l |}
\hline & \\
Spectral Type & G5 V \\
$V$\tablenotemark2 & $8.006 \pm 0.014$ \\
$B-V$\tablenotemark2 & $0.695 \pm 0.02$ \\
$M_{V}$ & $4.63 \pm 0.03$ \\
Parallax\tablenotemark2 & $21.06 \pm 1.04$ mas \\
Distance & $47 \pm 0.3$ pc \\
T$_{eff}$ & $5760 \pm 100$ K \\
$\log g$ & $4.45 \pm 0.12$ \\
$[$Fe/H$]$ & $0.24 \pm 0.06$ \\
$\xi$ & $1.20 \pm 0.15$ km/s \\
Mass\tablenotemark3 & $1.02 M_{\odot}$ \\
Age\tablenotemark3 & 7.20 Gyr \\
$\log R'_{HK}$ & $-4.65 \pm 0.03$ \\
\hline
\end{tabular}
\caption{Stellar Properties for HD 204313}
\label{stellar}

\tablenotetext{2}{\citet{kharchenko09}}
\tablenotetext{3}{From \citet{casagrande11}, maximum likelihood estimate using Padova isochrones.}

\end{center}
\end{table}

\begin{table}
\begin{center}

\begin{tabular}{| l l l |}
\hline & & \\
Orbital Parameter & Planet b & Planet d \\
\hline & & \\
Period $P$ (days) & $1920.1 \pm 25$ & $2831.6 \pm 150$ \\
Periastron Passage $T_0$ (BJD - 2 450 000) & $2111.6 \pm 28$ & $6376.9 \pm 176$ \\
RV Amplitude $K$ (m/s) & $57.0 \pm 3$ & $23.7 \pm 4$ \\
Mean Anomaly $M_0$\tablenotemark4 & $300^{\circ} \pm 0.4^{\circ}$ & $137^{\circ} \pm 2^{\circ}$ \\
Eccentricity $e$ & $0.23 \pm 0.04$ & $0.28 \pm 0.09$ \\
Longitude of Periastron $\omega$ & $298^{\circ} \pm 6^{\circ}$ & $247^{\circ} \pm 16^{\circ}$ \\
Semimajor Axis $a$ (AU) & $3.04 \pm 0.06$ & $3.93 \pm 0.14$ \\
Minimum Mass $M \sin i$ ($M_{J}$) & $3.55 \pm 0.2$ & $1.68 \pm 0.3$ \\
CORALIE RV offset (m/s) & \multicolumn{2}{c |}{-19.3} \\
2.7 m RV offset (m/s) & \multicolumn{2}{c |}{29.8} \\
RMS (m/s) & \multicolumn{2}{c |}{7.80} \\
Stellar ``jitter'' (m/s) & \multicolumn{2}{c |}{5.46} \\
\hline
\end{tabular}
\caption{Two-planet orbital solution for the HD 204313 system}
\label{orbit}
\tablenotetext{4}{Evaluated at the time of the first RV point reported in \citet{segransan10}}
\end{center}
\end{table}

\end{document}